%%POST REFEREE, Masiero ref added
\input harvmac
% +--------------------------------------------------------------------+
% |                                                                    |
% |                           TABLES.TEX                               |
% |                                                                    |
% |                     Ray F. Cowan  15-Feb-85                        |
% |                                                                    |
% |                       Princeton University                         |
% |                                                                    |
% |          Present Address:  Laboratory for Nuclear Science          |
% |                            M.I.T.                                  |
% |                            Cambridge, MA 02139                     |
% |                                                                    |
% |                   E-mail:  rfc@slacvm.slac.stanford.edu            |
% |                                                                    |
% |                                                                    |
% |                     Last Revision: 17-Apr-86                       |
% |                                                                    |
% |   Macros I find handy for making tables.  See TABLEDOC TEX for     |
% |   a longer description.  The token-counting macros are straight    |
% |   from the TeXbook's "Dirty Tricks" appendix.                      |
% |                                                                    |
% +--------------------------------------------------------------------+
%
\newbox\hdbox%
\newcount\hdrows%
\newcount\multispancount%
\newcount\ncase%
\newcount\ncols% This is the number of primary text columns in the table.
\newcount\nrows%
\newcount\nspan%
\newcount\ntemp%
\newdimen\hdsize%
\newdimen\newhdsize%
\newdimen\parasize%
\newdimen\spreadwidth%
\newdimen\thicksize%
\newdimen\thinsize%
\newdimen\tablewidth%
\newif\ifcentertables%
\newif\ifendsize%
\newif\iffirstrow%
\newif\iftableinfo%
\newtoks\dbt%
\newtoks\hdtks%
\newtoks\savetks%
\newtoks\tableLETtokens%
\newtoks\tabletokens%
\newtoks\widthspec%
%
%  Book-keeping stuff--see how often these macros are called.
%
%  MOD RFC 900221.
%  Removed usage logging:  it's too complicated under VM/XA.
%\immediate\write15{%
%CP SMSG GJMSINK TEXTABLE --> TABLE MACROS V. 851121 JOB = \jobname%
%}%
%
%  Turn on table diagnostics.
%
\tableinfotrue%
\catcode`\@=11%  Allows use of "@" in macro names, like PLAIN.TEX does.
%  Debugging aid.  Writes #1 on the
%                                    user's terminal and in the log file.
%
%  Define the \tstrut height, depth in terms of the x_height parameter.
%
\def\tstrut{\vrule height3.1ex depth1.2ex width0pt}%
\def\and{\char`\&}%  Allows us to get an `&' in the text.  This is the
%                    same as using the PLAIN TeX macro \&.
\def\tablerule{\noalign{\hrule height\thinsize depth0pt}}%
\thicksize=1.5pt%  Default thickness for fat rules.  The user should feel
%                  free to change this to his preference.
\thinsize=0.6pt%   Default thickness for thin rules.
\def\thickrule{\noalign{\hrule height\thicksize depth0pt}}%
\def\ctr#1{\hfil\ #1\hfil}%
%
%
%
%  Here are things for controlling the width of the finished table.
%
\tablewidth=-\maxdimen%
\spreadwidth=-\maxdimen%
\def\tabskipglue{0pt plus 1fil minus 1fil}%
%
%  Stuff for centering or not.
%
\centertablestrue%
%
%
%
%  \vctr vertically centers its argument in the row.
%
\parasize=4in%
\gdef\ARGS{########}%  Produces the correct number of #'s in the preamble
%                      by the time everything is expanded and \halign sees
%                      it.
\gdef\headerARGS{####}%  Same as \ARGS, but used in \header macros.
\def\@mpersand{&}%  Allows us to get alignment tab characters later
%                   when we have made the character "&" an active macro.
{\catcode`\|=13%  Make |'s locally active.
\gdef\letbarzero{\let|0}%  Globally define a macro that allows us to
%                          keep active |'s from being expanded in edef's.
\gdef\letbartab{\def|{&&}}%
\gdef\letvbbar{\let\vb|}%
%  This \def will cause active |'s read by
%                            \ruledtable to be converted into double
%                            alignment tabs.
}%  End of locally active |'s.
{\catcode`\&=4%  Make these alignment tabs.
\def\ampskip{&\omit\hfil&}%  This local macro skips a vertical rule.
\catcode`\&=13%  Now make &'s into active macros.
\let&0%  This allows us to expand \ampskip in the next \xdef without
%        attempting to expand the & and getting an "undefined control
%        sequence" error.
\xdef\letampskip{\def&{\ampskip}}%
\gdef\letnovbamp{\let\novb&\let\tab&}
%  This will cause active &'s read by
%                                   \ruledtable to be converted into
%                                   double tabs and an \omit'ted \vrule.
}%  End of locally active &'s.
\def\begintable{%  Here we make |'s and &'s active characters so we can
%                  interpret them as macros.  Note that this action is
%                  true only until we encounter the matching \endgroup
%                  token later at the end of the \ruledtable macro.
   \begingroup%
   \catcode`\|=13\letbartab\letvbbar%
   \catcode`\&=13\letampskip\letnovbamp%
   \def\multispan##1{%  We must redefine \multispan to count the number
%                       of primary columns, not physical columns.
      \omit \mscount##1%
      \multiply\mscount\tw@\advance\mscount\m@ne%
      \loop\ifnum\mscount>\@ne \sp@n\repeat%
   }%  End of \multispan macro.
   \def\|{%
      &\omit\widevline&%
   }%
   \ruledtable%  Now we call \ruledtable to do the real work.
}%  End of \begintable macro.
\long\def\ruledtable#1\endtable{%
%
%  This macro reads in the user's data entries
%  and converts them into a ruled table.
%
%  Important note:  Many macros and parameters are re-defined here, and
%  these must be kept local to the table macros to avoid conflict with
%  their use outside of tables.  This is done by the \begingroup token
%  macro \begintable and the \endgroup token at the end of
%  this macro.
%
   \offinterlineskip%  Needed to make rules touch each other.
   \tabskip 0pt%  Needed for same reason as \offinterlineskip.
   \def\widevline{\vrule width\thicksize}%  Make outer \vrule's wider.
   \def\endrow{\@mpersand\omit\hfil\crnorm\@mpersand}%
   \def\crthick{\@mpersand\crnorm\thickrule\@mpersand}%
   \def\crthickneg##1{\@mpersand\crnorm\thickrule
          \noalign{{\skip0=##1\vskip-\skip0}}\@mpersand}%
   \def\crnorule{\@mpersand\crnorm\@mpersand}%
   \def\crnoruleneg##1{\@mpersand\crnorm
          \noalign{{\skip0=##1\vskip-\skip0}}\@mpersand}%
   \let\nr=\crnorule%  A shorter abbreviation.
   \def\endtable{\@mpersand\crnorm\thickrule}%
   \let\crnorm=\cr%  Allows us to use \cr for our own purposes.
%
%  Cause user-typed \cr's to follow a row with a \tablerule.
%
   \edef\cr{\@mpersand\crnorm\tablerule\@mpersand}%
   \def\crneg##1{\@mpersand\crnorm\tablerule
          \noalign{{\skip0=##1\vskip-\skip0}}\@mpersand}%
   \let\ctneg=\crthickneg
   \let\nrneg=\crnoruleneg
   \the\tableLETtokens%  Get the user's extra \let's, if any.
%
%  Put the data entries into a token register so we can scan through them
%  and see what the user is asking us to do.
%
   \tabletokens={&#1}%  We add an extra alignment tab to the beginning
%                       of the first row to allow for the first \vrule.
%
%  Now count how many rows are in the table and return the result in
%  count register \nrows; do the same for columns, and return that
%  in register \ncols.
%
   \countROWS\tabletokens\into\nrows%
   \countCOLS\tabletokens\into\ncols%
%
%  Now do a little arithmetic to convert the number of primary columns
%  into the number of physical columns that the alignment preamble must
%  prepare for;  similarly for rows.
%
   \advance\ncols by -1%
   \divide\ncols by 2%
   \advance\nrows by 1%
%
%  Tell the user how many rows and columns we found in his data, if he
%  wants to know.
%
   \iftableinfo %
      \immediate\write16{[Nrows=\the\nrows, Ncols=\the\ncols]}%
   \fi%
%
%  Now we actually go ahead and produce the table.
%
   \ifcentertables
      \ifhmode \par\fi%  Make sure we are in vertical mode.
      \line{%  The final table comes out as an \hbox of width the \hsize.
      \hss%  The final table will be centered left-to-right.
   \else %
      \hbox{%
   \fi
      \vbox{%
         \makePREAMBLE{\the\ncols}%  Generate the preamble.
         \edef\next{\preamble}%  This line and the next line force the
         \let\preamble=\next%    expansion of all \ARGS tokens into the
%                                appropriate number of #'s.
         \makeTABLE{\preamble}{\tabletokens}%  Go do the \halign here.
      }%  End of \vbox.
      \ifcentertables \hss}\else }\fi%  Finish the centering effect.
%                                       It is important that no spaces
%                                       follow the two `}' here.
%  }%  End of \line.
   \endgroup%  Return all local macros and parameters to their outside
%              values.
   \tablewidth=-\maxdimen%  Reset \tablewidth to normal.
   \spreadwidth=-\maxdimen% Same for \spreadwidth.
}%  End of macro \ruledtable.
\def\makeTABLE#1#2{%  Does an \halign for the \ruledtable macro.
   {%  Start of local parameter values.
   \let\ifmath0%     These macros would cause trouble if they were to be
   \let\header0%     expanded in the following \xdef; we \let them be
   \let\multispan0%  equal to a digit, because digits can't be expanded.
%
%  Set up the width specification here.
%
   \ncase=0%
   \ifdim\tablewidth>-\maxdimen \ncase=1\fi%
   \ifdim\spreadwidth>-\maxdimen \ncase=2\fi%
   \relax%  This \relax is absolutely necessary, without it the following
%           \ifcase will always take \ncase=0.
%
   \ifcase\ncase %
      \widthspec={}%
   \or %
      \widthspec=\expandafter{\expandafter t\expandafter o%
                 \the\tablewidth}%
   \else %
      \widthspec=\expandafter{\expandafter s\expandafter p\expandafter r%
                 \expandafter e\expandafter a\expandafter d%
                 \the\spreadwidth}%
   \fi %
%\out{Widthspec=[\the\widthspec]}%
%\out{Preamble=[\preamble]}%
   \xdef\next{%  We must force the preamble to be expanded BEFORE the
      \halign\the\widthspec{%
%        \halign is done;  this \edef\next{...}\next construction
%                does the trick.
      #1%  This is the preamble text.
      \noalign{\hrule height\thicksize depth0pt}%  Makes the top \hrule.
      \the#2\endtable%  This is the main body.
%
%     \noalign{\hrule height0.7pt depth0pt}%  Makes the last \hrule.
      }%  End of \halign.
   }%  End of \next.
   }%  End of local values.
   \next%  This \next must be outside of the local values, because now
%          we want those troublesome macros in the \let's above to have
%          their normal actions.
}%  End of macro \makeTABLE.
\def\makePREAMBLE#1{%  This macro generates the necessary preamble for a
%                      ruled table with #1 primary columns.
%                      (Primary columns means the number of columns NOT
%                       counting those used for vertical rules.)
   \ncols=#1%  Get the number of columns desired.
   \begingroup%  Start local parameter definitions.
   \let\ARGS=0%  This is the key to the whole thing; it prevents \ARGS
%                from being expanded in the following \edef's.
   \edef\xtp{\widevline\ARGS\tabskip\tabskipglue%
   &\ctr{\ARGS}\tstrut}%  A 1-column preamble.  Gets the sizing right.
   \advance\ncols by -1%  One column has been generated; decrement the
%                         counter.
   \loop%  Append as many further columns as needed to the preamble.
      \ifnum\ncols>0 %
      \advance\ncols by -1%
      \edef\xtp{\xtp&\vrule width\thinsize\ARGS&\ctr{\ARGS}}%
   \repeat
   \xdef\preamble{\xtp&\widevline\ARGS\tabskip0pt%
   \crnorm}%  Adds the last \vrule.
   \endgroup%  End of local parameters.
}%  End of macro \makePREAMBLE.
\def\countROWS#1\into#2{%  This counts the number of rows in #1 by
%                          looking for control sequences that end a row,
%                          e.g., \cr, \crthick, etc., and puts the result
%                          into count register #2.
   \let\countREGISTER=#2%
   \countREGISTER=0%
%  \out{In countROWS:  tokens are [\the#1]}%
   \expandafter\ROWcount\the#1\endcount%
}%
\def\ROWcount{%
   \afterassignment\subROWcount\let\next= %
}%
\def\subROWcount{%
%  \out{In subROWcount:  next is [\meaning\next]}%  Debugging aid.
   \ifx\next\endcount %
      \let\next=\relax%
   \else%
      \ncase=0%
      \ifx\next\cr %
         \global\advance\countREGISTER by 1%
         \ncase=0%
      \fi%
      \ifx\next\endrow %
         \global\advance\countREGISTER by 1%
         \ncase=0%
      \fi%
      \ifx\next\crthick %
         \global\advance\countREGISTER by 1%
         \ncase=0%
      \fi%
      \ifx\next\crnorule %
         \global\advance\countREGISTER by 1%
         \ncase=0%
      \fi%
      \ifx\next\crthickneg %
         \global\advance\countREGISTER by 1%
         \ncase=0%
      \fi%
      \ifx\next\crnoruleneg %
         \global\advance\countREGISTER by 1%
         \ncase=0%
      \fi%
      \ifx\next\crneg %
         \global\advance\countREGISTER by 1%
         \ncase=0%
      \fi%
      \ifx\next\header %
%     \out{In subROWcount:  next=header, ncase set=1}%
         \ncase=1%
      \fi%
%     \out{In subROWcount:  ncase is [\the\ncase]}%
      \relax%
      \ifcase\ncase %
         \let\next\ROWcount%
%        \out{subROWcount---> ncase=\the\ncase}%
      \or %
         \let\next\argROWskip%
%        \out{subROWcount---> ncase=\the\ncase}%
      \else %
      \fi%
   \fi%
%  \out{subROWcount---> NEXT=\meaning\next}%
   \next%
}%  End of macro \subROWcount.
\def\counthdROWS#1\into#2{%
\dvr{10}%
   \let\countREGISTER=#2%
   \countREGISTER=0%
\dvr{11}%
%  \out{In counthdROWS:  tokens are [\the#1]}%
\dvr{13}%
   \expandafter\hdROWcount\the#1\endcount%
\dvr{12}%
}%
\def\hdROWcount{%
   \afterassignment\subhdROWcount\let\next= %
}%
\def\subhdROWcount{%
%\out{In subhdROWcount:  next is [\meaning\next]}%
   \ifx\next\endcount %
      \let\next=\relax%
   \else%
      \ncase=0%
      \ifx\next\cr %
         \global\advance\countREGISTER by 1%
         \ncase=0%
      \fi%
      \ifx\next\endrow %
         \global\advance\countREGISTER by 1%
         \ncase=0%
      \fi%
      \ifx\next\crthick %
         \global\advance\countREGISTER by 1%
         \ncase=0%
      \fi%
      \ifx\next\crnorule %
         \global\advance\countREGISTER by 1%
         \ncase=0%
      \fi%
      \ifx\next\header %
%\out{In subhdROWcount:  next=header, ncase set=1}%
         \ncase=1%
      \fi%
%\out{In subhdROWcount:  ncase is [\the\ncase]}%
\relax%
      \ifcase\ncase %
         \let\next\hdROWcount%
%\out{subhdROWcount---> ncase=\the\ncase}%
      \or%
         \let\next\arghdROWskip%
%\out{subhdROWcount---> ncase=\the\ncase}%
      \else %
      \fi%
   \fi%
%\out{subhdROWcount---> NEXT=\meaning\next}%
   \next%
}%
{\catcode`\|=13\letbartab
\gdef\countCOLS#1\into#2{%
%  \out{In countCOLS:  tokens are [\the#1]}
   \let\countREGISTER=#2%
   \global\countREGISTER=0%
   \global\multispancount=0%
   \global\firstrowtrue
   \expandafter\COLcount\the#1\endcount%
   \global\advance\countREGISTER by 3%
   \global\advance\countREGISTER by -\multispancount
%  \out{countCOLS-->[\the\countREGISTER]}
}%
\gdef\COLcount{%
   \afterassignment\subCOLcount\let\next= %
}%
{\catcode`\&=13%
\gdef\subCOLcount{%
%\out{In subCOLcount: next is [\meaning\next]}
   \ifx\next\endcount %
      \let\next=\relax%
   \else%
      \ncase=0%
      \iffirstrow
         \ifx\next& %
            \global\advance\countREGISTER by 2%
            \ncase=0%
         \fi%
         \ifx\next\span %
            \global\advance\countREGISTER by 1%
            \ncase=0%
         \fi%
         \ifx\next| %
            \global\advance\countREGISTER by 2%
            \ncase=0%
         \fi
         \ifx\next\|
            \global\advance\countREGISTER by 2%
            \ncase=0%
         \fi
         \ifx\next\multispan
            \ncase=1%
            \global\advance\multispancount by 1%
         \fi
         \ifx\next\header
            \ncase=2%
         \fi
         \ifx\next\cr       \global\firstrowfalse \fi
         \ifx\next\endrow   \global\firstrowfalse \fi
         \ifx\next\crthick  \global\firstrowfalse \fi
         \ifx\next\crnorule \global\firstrowfalse \fi
         \ifx\next\crnoruleneg \global\firstrowfalse \fi
         \ifx\next\crthickneg  \global\firstrowfalse \fi
         \ifx\next\crneg       \global\firstrowfalse \fi
      \fi%  End of \iffirstrow.
\relax%\out{subCOL-->  ncase=[\the\ncase]}
% \out{subCOL-->  next=\meaning\next}
      \ifcase\ncase %
         \let\next\COLcount%
      \or %
         \let\next\spancount%
      \or %
         \let\next\argCOLskip%
      \else %
      \fi %
   \fi%
%  \out{subCOL-->  countREGISTER=[\the\countREGISTER]}
   \next%
}%
\gdef\argROWskip#1{%
%  Deletes the next balanced, undelimited argument from a
%                 token list.
% \out{---> Entering argROWskip <---}
% \out{In argROWskip:  deleted arg is [#1]}%
   \let\next\ROWcount \next%
}%  End of macro \argskip.
\gdef\arghdROWskip#1{%
%  Deletes the next balanced, undelimited argument from a
%                 token list.
% \out{---> Entering arghdROWskip <---}
% \out{In arghdROWskip:  deleted arg is [#1]}%
   \let\next\ROWcount \next%
}%  End of macro \arghdROWskip.
\gdef\argCOLskip#1{%
%  Deletes the next balanced, undelimited argument from a
%                 token list.
% \out{---> Entering argCOLskip <---}
% \out{In argCOLskip:  deleted arg is [#1]}%
   \let\next\COLcount \next%
}%  End of macro \argskip.
}%  End of active &'s.
}%  End of active |'s.
\def\spancount#1{%\out{spancount--->\meaning#1}
   \nspan=#1\multiply\nspan by 2\advance\nspan by -1%
   \global\advance \countREGISTER by \nspan
%  \out{number spancount--->\the\nspan; \the\countREGISTER}
   \let\next\COLcount \next}%
\def\dvr#1{\relax}%
% \omit\hfil%
% \parindent=0pt\hsize=1.1in\valign{%
% \vfil#\vfil&\vfil#\vfil\cr\hfil\hbox{\ Added to\ }\hfil&%
% \hfil\hbox{\ empty events\ }\hfil\cr}\hfil%
\def\header#1{%
\dvr{1}{\let\cr=\@mpersand%
\hdtks={#1}%
%\out{In header:  hdtks=[\the\hdtks]}%
\counthdROWS\hdtks\into\hdrows%
\advance\hdrows by 1%
\ifnum\hdrows=0 \hdrows=1 \fi%
%\out{In header:  Nhdrows=[\the\hdrows]}%
\dvr{5}\makehdPREAMBLE{\the\hdrows}%
%\out{In header:  headerpreamble=[\headerpreamble]}%
\dvr{6}\getHDdimen{#1}%
%\out{In header:  hdsize=[\the\hdsize]}%
%\striplastCR{#1}%
{\parindent=0pt\hsize=\hdsize{\let\ifmath0%
\xdef\next{\valign{\headerpreamble #1\crnorm}}}\dvr{7}\next\dvr{8}%
}%
}\dvr{2}}%  End of macro \header.
\def\makehdPREAMBLE#1{%This macro generates the necessary preamble for a
\dvr{3}%
%                      ruled table with \ncols primary columns.
%                      (Primary columns means the number of columns NOT
%                       counting those used for vertical rules.
\hdrows=#1%  Get the number of columns desired.
{%  Start local parameter definitions.
\let\headerARGS=0%
%  This is the key to the whole thing; it prevents \ARGS
\let\cr=\crnorm%
%                from being expanded in the followin \edef's.
\edef\xtp{\vfil\hfil\hbox{\headerARGS}\hfil\vfil}%
\advance\hdrows by -1%  One row has been generated; decrement the
%                         counter.
\loop%  Append as many further rows as needed to the preamble.
\ifnum\hdrows>0%
\advance\hdrows by -1%
\edef\xtp{\xtp&\vfil\hfil\hbox{\headerARGS}\hfil\vfil}%
\repeat%
\xdef\headerpreamble{\xtp\crcr}%
}%  End of local parameters.
\dvr{4}}%  End of \makehdPREAMBLE.
\def\getHDdimen#1{%
%\out{In getHDdimen:  Arg 1=[#1]}%
\hdsize=0pt%
\getsize#1\cr\end\cr%
}%  End of macro getHDdimen.
\def\getsize#1\cr{%
%\out{In getsize:  Arg 1=[#1]}%
%  Here we have to check arg#1 and see if the first token in #1 is an
%    \end; if so, we stop, else we check the width of arg#1.
%  We recall that each arg#1 will be terminated with a \cr token.
\endsizefalse\savetks={#1}%
%\out{In getsize:  the savetks = [\the\savetks]}%
\expandafter\lookend\the\savetks\cr%
%\out{In getsize:  ifendsize = [\meaning\ifendsize]}%
\relax \ifendsize \let\next\relax \else%
\setbox\hdbox=\hbox{#1}\newhdsize=1.0\wd\hdbox%
\ifdim\newhdsize>\hdsize \hdsize=\newhdsize \fi%
%\out{In getsize:  hdsize=[\the\hdsize]}%
%\out{In getsize:  newhdsize=[\the\newhdsize]}%
\let\next\getsize \fi%
\next%
}%
\def\lookend{\afterassignment\sublookend\let\looknext= }%
\def\sublookend{\relax%
%\out{In sublookend:  looknext = [\looknext]}%
\ifx\looknext\cr %
%\out{In sublooknext:  looknext=cr}%
\let\looknext\relax \else %
%\out{In sublooknext:  looknext/=cr}%
   \relax
   \ifx\looknext\end \global\endsizetrue \fi%
   \let\looknext=\lookend%
    \fi \looknext%
}%
%
%  Allow the user to make his own names for crthick, etc.
%
\def\tablelet#1{%
   \tableLETtokens=\expandafter{\the\tableLETtokens #1}%
}%
\catcode`\@=12%  Change @'s back to their normal category code.
%
%%%%%%%%%%%%%%%%%%%%%%%%%%%%%%%%%%%
%%author definitions
\def \inparg{\leftskip = 40 pt\rightskip = 40pt}
\def \outparg{\leftskip = 0 pt\rightskip = 0pt}

\thicksize=0.7pt
\thinsize=0.5pt
\def\ctr#1{\hfil $\,\,\,#1\,\,\,$ \hfil}
\def\tstrut{\vrule height 2.7ex depth 1.0ex width 0pt}

\def\pa{\partial}
\def\Rcal{{\cal R}}

\def\semi{;\hfil\break}

\def\Ytilde{\tilde Y}

\def\frak#1#2{{\textstyle{{#1}\over{#2}}}}
\def\frakk#1#2{{{#1}\over{#2}}}
\def\mbar{{\overline{m}}}
\def\qbar{{\overline{q}}}
\def\Kbar{{\overline{K}}}
\def\kappabar{{\overline{\kappa}}}

\def\semi{;\hfil\break}
\def\npb{{Nucl.\ Phys.\ }{\bf B}}
\def\prd{{Phys.\ Rev.\ }{\bf D}}

\def\plb{{Phys.\ Lett.\ }{\bf B}}

\def\abar{{\overline a}}
\def\thetabar{{\overline \theta}}

\def\Rcal{{\cal R}}  

\def\ttil{\tilde t}
\def\btil{\tilde b}
\def\tautil{\tilde \tau}
\def\util{\tilde u}
\def\dtil{\tilde d}
\def\etil{\tilde e}

\def\nutil{\tilde \nu} 
\def\chitil{\tilde \chi} 
\def\TeV{{\rm TeV}}
\def\GeV{{\rm GeV}}
{\nopagenumbers
\line{\hfil LTH 479}
\line{\hfil hep-ph/0006116}
\vskip .5in
\centerline{\titlefont R-symmetry, Yukawa textures  and } 
\centerline{\titlefont anomaly mediated supersymmetry breaking}
\vskip 1in
\centerline{\bf I.~Jack and D.R.T.~Jones}
\bigskip
\centerline{\it Dept. of Mathematical Sciences,
University of Liverpool, Liverpool L69 3BX, U.K.}
\vskip .3in

We explore, in the MSSM context, an extension of  the  Anomaly Mediated
Supersymmetry Breaking solution for the  soft scalar masses  that is
possible if the underlying theory has  a gauged R-symmetry. The 
slepton mass problem characteristic of the scenario is resolved,
and a context for the explanation of the fermion mass  hierarchy provided.

\Date{June 2000}}

%\newsec{Introduction}
Recently there has been interest in a 
specific and  predictive  framework for the origin of soft supersymmetry 
breaking within the MSSM, known as 
Anomaly Mediated Supersymmetry Breaking (AMSB).
The supersymmetry-breaking terms originate in a
vacuum expectation value for an F-term in the  supergravity
multiplet, and the 
gaugino mass $M$, the $\phi^3$ coupling
$h^{ijk}$ and the  $\phi\phi^*$-mass $(m^2)^i{}_j$ are all given in terms  of
the gravitino mass, $m_0$,  and the $\beta$-functions of the
unbroken theory  by simple relations that are renormalisation group
(RG) invariant
\ref\lisa{L. Randall and R. Sundrum, \npb 557 (1999) 79}
\nref\glmr{G.F. Giudice, M.A. Luty, H. Murayama and  R. Rattazzi,
JHEP 9812 (1998) 27}%
\nref\jjpa{I.~Jack, D.R.T.~Jones and A.~Pickering,
\plb426 (1998) 73}%
\nref\kkz{T.~Kobayashi, J.~Kubo and G.~Zoupanos, \plb427 (1998) 291}%
\nref\appp{A. Pomarol and  R. Rattazzi, JHEP 9905 (1999) 013}%
\nref\ggw{T. Gherghetta, G.F. Giudice and J.D. Wells, \npb 559 (1999) 27}%
\nref\clmp{Z. Chacko, M.A. Luty, I. Maksymyk and E. Ponton, 
JHEP 0004 (2000) 001}%
\nref\lura{M.A. Luty and R. Rattazzi, JHEP 9911 (1999) 001}% 
\nref\kss{E. Katz, Y. Shadmi and Y. Shirman, JHEP 9908 (1999) 015}% 
\nref\jjrg{I.~Jack and D.R.T.~Jones, \plb 465 (1999) 148}%
\nref\jlftm{J.L.~Feng and T.~Moroi, \prd 61 (2000) 095004}%
\nref\gdk{G.D.~Kribs, \prd 62 (2000) 015008}%
\nref\shusu{S.~Su, \npb 573 (2000) 87}%
\nref\rzzsw{R. Rattazzi, A. Strumia and J.D. Wells, hep-ph/9912390}%
\nref\fepjw{F.E.~Paige and J. Wells, hep-ph/0001249}%
\nref\jjnew{I.~Jack and D.R.T~Jones, hep-ph/0003081}%
\nref\marcel{M. Carena, K. Huitu and T. Kobayashi, hep-ph/0003187}%
\nref\allanach{B.C. Allanach and A. Dedes, hep-ph/0003222}%
\nref\clppss{Z. Chacko et al, hep-ph/0006047}%
--\ref\chghro{U. Chattopadhyay, D.K. Ghosh and S. Roy, hep-ph/0006049}.
Direct application of this idea to the MSSM leads, unfortunately, 
to negative $(\hbox{mass})^2$ sleptons: in other words, to a theory 
without a vacuum 
preserving the $U_1$ of electromagnetism. Various resolutions of this dilemma 
have been investigated; here we explore a particularly minimalist one, 
which requires the introduction of no new fields into the low energy 
theory.    The key lies in a compelling generalisation of the RG 
invariant solution described above\appp. The basic AMSB solution is 
given by:
\eqna\result$$\eqalignno{M &= m_0{\beta_g\over g}, &\result a\cr
h^{ijk}&=-m_0\beta_Y^{ijk},&\result b\cr
%b^{ij}&=-m_0\beta_{\mu}^{ij}, &\result c\cr
(m^2)^i{}_j &= \frak{1}{2}|m_0|^2\mu\frakk{d\gamma^i{}_j}{d\mu}.
&\result c\cr}$$
Now $\beta_{m^2}$ is given by\jjpa\ (see also
\ref\yyamada{Y. Yamada, \prd 50 (1994) 3537}%
\nref\guiratt{G.F. Giudice and R. Rattazzi \npb 511 (1998) 25}% 
\nref\arkham{N.~Arkani-Hamed  et al, \prd 58 (1998) 115005}%
\nref\akk{L.V.~Avdeev, D.I.~Kazakov and
I.N.~Kondrashuk, \npb510 (1998) 289}%
--\ref\newkaz{D.I.~Kazakov and V.N.~Velizhanin, 
hep-ph/0005185})
\eqn\Ajy{
(\beta_{m^2})^i{}_j (m^2, \cdots) =\left[
2{\cal O}{\cal O}^* +2MM^{*} g^2{\pa
\over{\pa g^2}} +\Ytilde{\pa\over{\pa Y}}
+\Ytilde^{*}{\pa\over{\pa Y^{*}}} + 
X\frakk{\pa}{\pa g}\right]
\gamma^i{}_j,}
where
\eqn\Ajb{
{\cal O}=\left(Mg^2{\pa\over{\pa g^2}}-h^{lmn}{\pa
\over{\pa Y^{lmn}}}\right),}
\eqn\Ajd{
\Ytilde^{ijk}(m^2,Y) =(m^2)^i{}_lY^{ljk}+(m^2)^j{}_lY^{ilk}+(m^2)^k{}_lY^{ijl}}
and (in the NSVZ scheme)\kkz\ref\jjpb{I.~Jack, D.R.T.~Jones and A.~Pickering,
\plb 432 (1998) 114}
\foot{In Ref.~\newkaz\  
the existence of $X$ (absent in Ref.~\akk)\ is confirmed} 
\eqn\exX{
X(m^2,M)  =-2{g^3\over{16\pi^2}}
{ {r^{-1}\Tr [m^2C(R)] -MM^* C(G)}
\over{\left[1-2g^2 C(G)(16\pi^2)^{-1}\right]}}.}
(Here $r$ is the number of generators of the gauge group and
$C(R)$ and $C(G)$ are the quadratic matter and adjoint Casimirs respectively.)

It is immediately clear that, given a solution to Eq.~\Ajy,  
$m^2= m^2_1$, then $m^2= m^2_1 + m^2_2$ is also a solution, 
where $m^2_2$ satisfies the equation (linear and homogeneous in $m_2^2$):
\eqn\Rsyma{\mu\frakk{d}{d\mu}m^2_2 = 
\left[\Ytilde^* (m_2^2,Y^*)\frakk{\pa}{\pa Y^*}
+\Ytilde (m_2^2,Y)\frakk{\pa}{\pa Y}
+ X(m^2_2, 0)\frakk{\pa}{\pa g}\right]\gamma.}
Remarkably, Eq.~\Rsyma\ has a solution of the form\appp\lura 
\eqn\Rsol{
(m_2^2)^i{}_j = \mbar_0^2 (\gamma^i{}_j + \qbar^i \delta^i{}_j)}
where $\mbar_0^2$ and $\qbar^i$ are constants, as long as 
a set $\qbar^i$ exists that satisfy the following constraints:
\eqna\qrels$$\eqalignno{
(\qbar^i+\qbar^j+\qbar^k)Y_{ijk} &= 0 &\qrels a\cr
2\Tr\left[\qbar C(R)\right] + Q &= 0, & \qrels b\cr}$$
where $Q$ is the one loop $\beta_g$ coefficient. It is easy to show\appp\  
that Eq.~\qrels{}\ corresponds precisely to requiring 
that the theory have a non-anomalous  
R-symmetry (which we will denote ${\cal R}$, to 
avoid confusion with our notation $R$ for group representations). 
Setting
\eqn\qcharg{\qbar^i = 1 -\frak{3}{2}r^i,}
we see that Eq.~\qrels{a}\ corresponds to $(r^i+r^j+r^k)Y_{ijk} = 2Y_{ijk}$, 
which is the conventional ${\cal R}$-charge normalisation. Moreover, it is 
then easy to show (recall that the gaugino has $\Rcal$-charge of $1$) that
Eq.~\qrels{b}\ is simply the anomaly cancellation condition for the 
$\Rcal$-charges.     

Our strategy in this paper will be to take the AMSB solution Eq.~\result{}, 
but with Eq.~\result{c}\ generalised to 
\eqn\mnew{(m^2)^i{}_j = \frak{1}{2}|m_0|^2\mu\frakk{d\gamma^i{}_j}{d\mu}
+ \mbar_0^2 (\gamma^i{}_j + \qbar^i \delta^i{}_j).}
For a discussion of a possible origin of $\mbar_0^2$ as the vacuum 
expectation value of a $U_1$ $D$-term, see Ref.~\lura.

In a theory with direct product structure there is a relation of the
form Eq.~\qrels{b}\ for each gauged subgroup; so in the MSSM case there
are three conditions, corresponding to cancellation of the $\Rcal
(SU_3)^2$, $\Rcal (SU_2)^2$ and $\Rcal (U_1)^2$ anomalies. As discussed 
in \ref\iban{L.~Ibanez, \plb 303 (1993) 55}, these anomalies could be
cancelled by the Green-Schwarz  mechanism, but this would not be
appropriate for us here since in that case  Eq.~\qrels{b} would no
longer be satisfied. We also impose cancellation of the $(\Rcal)^2 U_1$
anomaly, although this is not required to render  Eq.~\mnew\
RG-invariant. Cancellation of the remaining $(\Rcal)^3$ and
$\Rcal$-gravitational anomalies can also be achieved if  we assume the
existence of a MSSM-singlet sector (at high energies)\foot{Note that the
 gravitino also contributes to these anomalies \ref\chamdr{A.H.
Chamseddine and H. Dreiner, \npb 458 (1996) 65} \ref\danf{D.J. Castano,
D.Z. Freedman and C. Manuel, \npb 461 (1996) 50}.}.

Now for the MSSM superpotential
\eqn\spt{
W_{MSSM} = \mu_{s} H_1 H_2 + (\lambda_u)_{ab} H_2
Q_a (u^c)_b + (\lambda_d)_{ab} H_1 Q_a (d^c)_b  
+ (\lambda_{e})_{ab}H_1 L_a (e^c)_b,}
there is no possible $\Rcal$-symmetry, satisfying the 
constraints described above, 
such that all the Yukawa couplings are 
non-zero\foot{Application of the scenario to the MSSM was dismissed 
in Ref.~\appp, presumably for this reason.}. One way out of this dilemma
would be to add extra particles\chamdr; here we instead persist 
with the minimal field content, and are hence forced to 
distinguish between 
the generations. Apart from simplicity this also provides a context for 
explaining the fermion mass hierarchy. We therefore presume an $\Rcal$-charge 
assignment such that only the third generation Yukawa couplings are 
permitted (we will return later to the origin of the first two generation 
masses). We will, however, enact the constraint that the first 
two generations have identical $\Rcal$-charges. 
As we shall see, this will alleviate potential  
Flavour Changing
Neutral Current (FCNC) problems.

Thus for the superpotential to have $\Rcal$-charge $2$, 
we require (henceforth we work with the fermionic charges,
related to the $\Rcal$-charges by $q_f = r -1$): 
\eqna\gainv$$\eqalignno{
q_3 + u_3 + h_2 = q_3 +d_3 + h_1 = l_3 + e_3 + h_1 & = -1 & \gainv a\cr
h_1 + h_2 & = 0, & \gainv b\cr}$$
while for cancellation of the mixed anomalies we require
\eqna\anomcanc$$\eqalignno{
q_3 +\frak{1}{2}(u_3 +d_3) + 2\left(q_1 
+\frak{1}{2}\left(u_1 +d_1\right)\right) + 3 &= 0 &\anomcanc a\cr
\frak{1}{2}l_3 + \frak{3}{2}q_3 + 2(\frak{1}{2}l_1 + \frak{3}{2}q_1)
+ \frak{1}{2}(h_1 + h_2) +2  &= 0 &\anomcanc b\cr
\frak{1}{6}q_3+\frak{1}{3}d_3+\frak{4}{3}u_3+\frak{1}{2}l_3 + e_3
+2(\frak{1}{6}q_1+\frak{1}{3}d_1+\frak{4}{3}u_1+\frak{1}{2}l_1 + e_1)
+ \frak{1}{2}(h_1 + h_2) 
&= 0  &\anomcanc c\cr
-l_3^2+e_3^2+q_3^2-2u_3^2+d_3^2
+2(-l_1^2+e_1^2+q_1^2-2u_1^2+d_1^2)-h_1^2+h_2^2 &= 0  &\anomcanc d\cr}$$
Eqs.~\anomcanc{a-d} correspond to cancellation of the 
$\Rcal (SU_3)^2$, $\Rcal (SU_2)^2$, $\Rcal (U_1)^2$ and $\Rcal^2 U_1$  
anomalies respectively. It is easy to show that 
even without imposing the quadratic constraint 
Eq.~\anomcanc{d}, the system of equations Eqs.~\gainv{}, 
\anomcanc{}\ has no solution if we set $q_1=q_3, u_1=u_3$ etc. 
Thus, as  asserted above, there is no possible generation independent 
$\Rcal$-charge assignment. 
The above constraints may be solved (for arbitrary values of the 
{\it leptonic\/} charges) as follows:
\eqna\ansol$$\eqalignno{
q_3 &= \frak{4}{9}-\frak{1}{3}l_3 -\frak{1}{9}\frak{\kappabar}{\kappa}
& \ansol a\cr
u_3 &= -\frak{22}{9} -\frak{2}{3}l_3 - e_3 
+\frak{1}{9}\frak{\kappabar}{\kappa}
& \ansol b \cr
d_3 &= -\frak{4}{9} +\frak{4}{3}l_3 + e_3 
+\frak{1}{9}\frak{\kappabar}{\kappa}
& \ansol c \cr
q_1 &= -\frak{101}{90} -\frak{1}{3}\kappa +\frak{1}{15}l_3 +\frak{1}{5}e_3 
+\frak{1}{30}\kappabar  +\frak{1}{18}\frak{\kappabar}{\kappa}
& \ansol d \cr
u_1 &= -\frak{79}{90} -\frak{2}{3}\kappa -\frak{16}{15}l_3 
-\frak{6}{5}e_3 
-\frak{1}{30}\kappabar - \frak{1}{18}\frak{\kappabar}{\kappa}
& \ansol e \cr
d_1 &= \frak{101}{90} +\frak{4}{3}\kappa +\frak{14}{15}l_3 
+\frak{4}{5}e_3-\frak{1}{30}\kappabar 
- \frak{1}{18}\frak{\kappabar}{\kappa}& \ansol f \cr
h_2 &= - h_1 = l_3+e_3 +1, & \ansol g \cr}$$
where $\kappa = l_1-l_3+e_1-e_3 - 3$, and 
$\kappabar= -12l_3-16e_3+10e_1 -23$. 
Thus for any set of rational values for the leptonic charges there 
exist rational values for all the charges.

We will presently exhibit a set of sum-rules for the sparticle masses 
that are completely independent of the set of values 
$l_3,e_3,\kappa,\kappabar$. 
Let us first see whether we can gain any insight on 
the $\Rcal$-charge assignments by relating them to  
a possible origin of the light quark and lepton masses. 
Suppose
\ref\cdfr{C.D. Froggatt and H.B. Nielsen, \npb 147 (1979) 277\semi
P. Ramond, R.G. Roberts and  G.G. Ross,
 \npb 406 (1993) 19\semi
L.E. Ib\'a\~nez and G.G. Ross, \plb 332 (1994) 100\semi 
P.~Binetruy and P.~Ramond, \plb 350 (1995)\semi  
E. Dudas, S. Pokorski and C.A. Savoy, \plb 356 (1995) 45} 
there are higher-dimension terms in the effective field 
theory of the form (for the up-type quarks)
$H_2 Q_i u^c_j (\frakk{\theta}{M_U})^{a_{ij}}$ or
$H_2 Q_i u^c_j (\frakk{\thetabar}{M_U})^{\abar_{ij}}$, 
where $\theta,\thetabar$ is a pair of 
MSSM singlet fields with $\Rcal$-charges $\pm r_{\theta}$ 
that get equal vacuum expectation values, 
and $M_U$ represents some high energy new physics scale (with similar 
terms for the light down quarks and leptons). Evidently the $\Rcal$-charge 
assignments will then dictate the texture of the Yukawa couplings, 
via the relation $h_2 +q_1 +u_1 +a_{11} r_{\theta}= -1$ and similar identities.

We thus obtain 
Yukawa textures of the general form:
\eqn\textureq{\Delta_u = 
\pmatrix{\epsilon^{\sigma| \kappa|} 
& \epsilon^{\sigma| \kappa|} & \epsilon^{\sigma| \delta_q|}\cr
\epsilon^{\sigma| \kappa|} 
& \epsilon^{\sigma| \kappa|} & \epsilon^{\sigma| \delta_q|}\cr
\epsilon^{\sigma| \kappa+\delta_q|} 
& \epsilon^{\sigma| \kappa+\delta_q|} & 1\cr},\quad
\Delta_d =
\pmatrix{\epsilon^{\sigma| \kappa|} 
& \epsilon^{\sigma| \kappa|} & \epsilon^{\sigma| \delta_q|}\cr
\epsilon^{\sigma| \kappa|} 
& \epsilon^{\sigma| \kappa|} & \epsilon^{\sigma| \delta_q|}\cr
\epsilon^{\sigma| \kappa-\delta_q|} 
& \epsilon^{\sigma| \kappa-\delta_q|} & 1\cr}}
for the up and down quarks, and 
\eqn\textua{\Delta_L =
\pmatrix{\epsilon^{\sigma| \kappa+3|} 
& \epsilon^{\sigma| \kappa+3|} & \epsilon^{\sigma| \delta_L|}\cr
\epsilon^{\sigma| \kappa+3|} 
& \epsilon^{\sigma| \kappa+3|} & \epsilon^{\sigma| \delta_L|}\cr
\epsilon^{\sigma| \kappa+3-\delta_L|} 
& \epsilon^{\sigma| \kappa+3-\delta_L|} & 1\cr}}
for the leptons, where
\eqn\deltdefs{\eqalign{
\delta_q &= \frakk{1}{30\kappa}(-47\kappa+12l_3 \kappa
+6e_3 \kappa+\kappa\kappabar -10\kappa^2+5\kappabar)\cr
\delta_L &=  \frakk{7}{10} -\frakk{6}{5}l_3-\frakk{3}{5}e_3
-\frakk{1}{10}\kappabar+\kappa,\cr}}
$\epsilon=\left|{<\theta>\over{M_U}}\right|$ 
and $\sigma= (|r_{\theta}|)^{-1}$ (provided $r_{\theta}$ is such that all
the exponents in Eqs.~\textureq, \textua\ are integers). More complex scenarios
may be contemplated in which there are more than one pairs of $\theta$, 
$\thetabar$ fields, but we do not consider this further.

In work on Yukawa textures
it is common to assume that they  
are symmetric: this assumption 
is not dictated by the theoretical structure of our model.
Moreover, it is 
easy to show  that to obtain symmetric textures for both up and down quarks 
requires $\kappa=\kappabar=0$. This then implies that the 
up and down quark Yukawa couplings amongst the 1st and 2nd generations are also 
allowed (and presumably of $O(1)$, leaving the fermion mass hierarchy 
unexplained). We therefore abandon the symmetric paradigm; as an alternative 
simplifying assumption, motivated by the similarity of the hierarchies of
the down quark and lepton masses, we impose $\Delta_d=\Delta_L$. 
This requires 
\eqn\dl{
\kappa=-\frak{3}{2}, \qquad \kappabar=-\frak{21}{2}-\frak{9}{4}\lambda,}
where $\lambda=2l_3+e_3$. We then find 
$\delta_q=\frak{3}{8}\lambda-\frak{1}{4}$. The only value of $\lambda$ we have
found which leads to nice textures with only one pair of $\theta$, 
$\thetabar$ fields is $\lambda=-\frak{1}{3}$; with $r_{\theta}=\frak{3}{8}$,
we then obtain texture matrices of the form
\eqn\textureq{\Delta_u =
\pmatrix{\epsilon^4
& \epsilon^4 & \epsilon\cr
\epsilon^4
& \epsilon^4 & \epsilon\cr
\epsilon^5
& \epsilon^5 & 1\cr},\quad
\Delta_d = \Delta_L=
\pmatrix{\epsilon^4
& \epsilon^4 & \epsilon\cr
\epsilon^4
& \epsilon^4 & \epsilon\cr
\epsilon^3
& \epsilon^3 & 1\cr}.}
 
The charges now have the form shown in Table~1.
\vskip3em
\vbox{
\begintable
 q_3      | l_3           | u_3     | d_3 | e_3
\cr
   \frak{e}{6} - \frak{2}{9}
  |-\frak{e}{2}-\frak{1}{6}|-\frak{2e}{3}-\frak{29}{18}
  |\frak{e}{3}+\frak{1}{18}  | e 
\endtable
%}
%and
\vskip1em
%\vbox{
\begintable
 q_1      | l_1           | u_1     | d_1 | e_1
| H_1      | H_2     \cr
   \frak{e}{6}-\frak{43}{72}|-\frak{e}{2}+\frak{5}{24}
  | -\frak{2e}{3}+\frak{19}{72}        | \frak{e}{3}-\frak{77}{72}
  |e+\frak{9}{8}            
  | -\frak{e}{2} -\frak{5}{6}| \frak{e}{2} +\frak{5}{6}
\endtable

\bigskip
\inparg
{\noindent {\it Table~1:\/} The fermionic $\Rcal$-charges 
for the case $\Delta_d=\Delta_L$}
\bigskip \outparg}

It is easy to show that  as long as $-\frak{1}{3} < e < \frak{1}{3}$ and
$\mbar_0^2<0$,
the contribution to each slepton mass term due to the $\qbar$ term in 
Eq.~\mnew\ will be positive, and we may expect to achieve a viable spectrum;
however, it turns out that it is still non-trivial to obtain  
an acceptable minimum because, for example, if $e=0$ and $\mbar_0^2<0$,
the $\mbar_0^2\qbar$ contributions to Eq.~\mnew\ from $u_3$, $q_1$ and $d_1$
are negative.  
Reverting to the Yukawa texture issue, 
we see that $\Delta_{u,d,L}$ are not in the class of forms for the texture 
matrix most 
frequently considered in the literature, where more attention 
has focussed on the possibility of texture zeroes.  They are of interest, 
however, in that $\Delta_u$ has one zero eigenvalue, and $\Delta_{d,L}$ have 
two zero eigenvalues. It follows from these properties that 
mass hierarchies may be produced with matrices of this generic structure. 
For example,
given the following up and down-quark Yukawa matrices,
\eqn\yukans{
\lambda_u \propto \pmatrix{-0.28\epsilon^4 & 1.3\epsilon^4 & 0.4\epsilon\cr
-0.32\epsilon^4 & 1.45\epsilon^4 & 1.36\epsilon\cr
-0.36\epsilon^5 & 1.67\epsilon^5 & 1\cr}\quad 
\lambda_d \propto 
\pmatrix{-1.75\epsilon^4 & 1.99\epsilon^4 & 0.25\epsilon\cr
-3.01\epsilon^4 & 2.53\epsilon^4 & 1.18\epsilon\cr
0.26\epsilon^3 & -0.48\epsilon^3 & 0.95\cr},} with $\epsilon = 0.25$,
we obtain ratios  for the quark masses and a CKM matrix within 
experimental limits\foot{We have neglected $CP$-violation, but  this can
easily be incorporated.}.  
Let us consider the issue of FCNC contributions (for a review, see 
for example Ref.~\ref\masier{A.~Masiero and L.~Silvestrini, 
hep-ph/9711401}). The matrices 
$\lambda_u$ and $\lambda_d$ are both diagonalised by matrices 
which are approximately of the 
general form
$$
\pmatrix{\cos\theta & \sin\theta & 0\cr
-\sin\theta & \cos\theta & 0\cr 
0 & 0 & 1\cr}
$$
from which it follows, because we chose identical $\Rcal$-charge 
assignments for the first two generations, that 
if we rotate the squark masses
to the basis that diagonalises both the quark masses and the 
quark-squark-gluino coupling, then all the off-diagonal terms are 
small, so FCNC contributions mediated by the gluino will be suppressed. 
Of course even in the absence of squark-flavour mixing there are susy 
FCNC contributions; consider for example the wino-squark box diagram 
contribution to $K-\Kbar$ mixing. Here the up/charm squark contributions will
be GIM suppressed and the top squark contribution suppressed by CKM angles, 
just as the analogous  Standard Model top quark diagram is. 
For the charged leptons,  we are less
constrained given the lack  of a (or, if we generalised to the massive
neutrino case, our ignorance  of the) leptonic CKM matrix. 

Naturally because the off-diagonal squark and slepton 
masses are (though 
relatively small) not zero, 
it follows that the whole issue of FCNCs deserves a more detailed 
analysis. 

We cannot entirely claim avoidance of fine-tuning, inasmuch as 
the lightest quark masses
($m_{u,d}$) are somewhat sensitive to small changes in the 
coefficients shown in Eq.~\yukans; for example if we change $1.3$ to $1.4$
in $\lambda_u$ then $m_u$ increases by a factor of $4$. 
However, the CKM matrix, $m_s$ and $m_c$ are remarkably stable 
under such variations.  

The mechanism proposed for generating the light fermion masses  raises
the following issue.  As  a symmetry of the low energy effective field
theory, our $\Rcal$-symmetry forbids from the superpotential, Eq.~\spt,
not only the light fermion Yukawa couplings but also the well-known set
of baryon and lepton-number violating terms of the  form $QLd^c$, $d^c
d^c u^c$, $LLe^c$ and $H_2 L$. It is clear that  {\it a
priori\/} the same mechanism we invoke above to generate the  light
masses might lead to similar contributions to these  operators, for
example via the  operator  $d^c_1 d^c_2 u^c_3
(\frakk{\theta}{M_U})^{p}.$ However  it is easy  to check that, with the
charge assignment we make above for the $\theta,
\thetabar$ fields, the value of $p$ required to render this operator 
$\Rcal$-invariant is not an integer; and similarly for the other
baryon and lepton-number violating operators above. There will in
general be higher-dimensional B-violating  and L-violating operators but
the effects of these will be strongly suppressed.

The phenomenology of AMSB-models has been discussed at length  in the
literature. If we compare our model here  with the constrained MSSM
(where the assumption of soft universality  at the unification scale
means that the theory is characterised by  the usual input parameters,
$\tan\beta$, $m_0$, $m_{\frak{1}{2}}$ and $A$), we see that  we have the
same number of parameters, $\tan\beta$,  $m_0$, $\mbar_0^2$ and the 
$\Rcal$-charge $e$.  We can try and further constrain the model by demanding
that the  soft $H_1H_2$ mass term lies on the same RG trajectory as the
other soft terms (see Ref.~\jjpa), but we find it impossible to find a
satisfactory vacuum in that case.

A characteristic feature of AMSB models is the near-degenerate light 
charged and neutral winos;  this prediction, depending as it does on
Eq.~\result{a},  is preserved in the scenario presented here.   A
variety of mass spectra for $m_0=40\TeV$ (corresponding to a gluino mass
of around $1\TeV$), but with different values of $\tan\beta$, $e$ and
$\mbar^2_0$, is  presented in Table 2;  we were unable to find any values 
of $e$ and $\mbar^2_0$ corresponding to an acceptable spectrum for 
$\tan\beta$ significantly larger than $10$.  
The heaviest sparticle masses scale with $m_0$ and are given roughly
by $M_{\rm SUSY} = \frak{1}{40}m_0$. Consequently we
take account of leading-log corrections by evaluating the mass
spectrum at this scale.
In other words, before applying Eq.~\mnew, 
we evolve the dimensionless couplings (together with $v_1$, $v_2$)
from the weak scale up to the scale $M_{\rm SUSY}$.
A dramatic feature of the 
spectra  is the splitting in the slepton masses for different 
generations.  Moreover, unusual\ref\hebb{T. Hebbeker, \plb 470 (1999)
259}\  is the possibility  (exemplified in the first three columns of
Table 2) that the  $\nutil_{\tau}$ is the LSP. 
As is well known, radiative corrections   give a
sizeable upward contribution  to the mass of the light CP-even Higgs, 
and so we have included the one-loop
calculation (in the  approximation given by Haber\ref\hrev{C.~Caso et
al,  The European Physical Journal (1998) 1}). 

\vfil\eject
\vbox{
\begintable
%\hbox {Particle} 
\tan\beta | 2  |2  | 5  |5 |10\cr 
\hbox{sign}~\mu_s | +  | - | +  |  + |+\cr
e| -1/9 | -1/9  | -1/9  | -2/9 |-2/9\cr
\mbar_0^2 (\TeV^2)|  -0.1|-0.1  | -0.1  |  -0.25|-0.2\cr
 \ttil_1 | 652 |615| 567  |302 |404\cr
\ttil_2 | 882| 908 |876  |879|875\cr   
\btil_1 | 865  |865 | 843  |853 |843\cr
\btil_2 | 977  | 977 |974  |1009 |987\cr
\tautil_1 | 94  | 87 | 75  | 136  |86\cr
\tautil_2 |110 | 116 |127  |289|251\cr
\util_L |918 |918|917  | 880  |892\cr
\util_R |997| 997|997  | 1084 |1057\cr
\dtil_L |920|920|921 | 884 |896\cr
\dtil_R |887|887|887 |776 |814\cr
\etil_L |260|260|261 |473 |418\cr
\etil_R |423|423|423 |664 |590\cr
\nutil_{\tau}| 83|83| 73 | 277 |234\cr 
\nutil_e | 251|251| 249 |467 |410\cr   
h | 96 | 105 |119  |114 |124\cr
H | 598 | 598 | 585  | 121 |308\cr
A | 593| 593|584  |110 |307\cr
H^{\pm} | 599|599| 590 |137 |318\cr  
\chitil^{\pm}_1 | 98 |116 | 104 | 101|106\cr
\chitil^{\pm}_2 | 628 |625 | 663 |449 |530\cr
\chitil_1 |98| 115 |103 |99|103\cr
\chitil_2 | 364 |372 | 367 |357|365 \cr
\chitil_3 | 619 | 620 |662  |446 |532\cr
\chitil_4 |637| 628 | 672  |470| 544 \cr
{\tilde g} | 1008 | 1008 |1008    | 1008|1008
 \endtable
\bigskip
\inparg
{\noindent {\it Table~2:\/} The sparticle masses (given in $\GeV$)}
\bigskip \outparg}

A salient feature of the model is the existence of sum rules in which the 
dependence on the $\Rcal$-charge assignment cancels. These sum rules 
follow from Eq.~\ansol{}; and thus for the particular solution 
exhibited in Table~1, they are independent of $e$. We find the
following relations for the physical masses (in each case independent of
$e$ and $\hbox{sign}~\mu_s$; in general the numerical results depend 
on $\tan\beta$, here taken throughout to equal 5, and also 
on $m_0$, here taken throughout to be $40\TeV$, 
due to the running to $M_{\rm SUSY}$ (which depends on $m_0$)):
\eqna\sumthr $$\eqalignno{
m_{\ttil_1}^2+m_{\ttil_2}^2+m_{\btil_1}^2+m_{\btil_2}^2
- 2(m_t^2 + m_b^2)-2.75m^2_{\tilde g}&= 0.92\mbar_0^2\TeV^2, & \sumthr a \cr
m_{\tautil_1}^2+m_{\tautil_2}^2+m_{\ttil_1}^2+m_{\ttil_2}^2
- 2(m_t^2 + m_{\tau}^2) -1.14m_{\tilde g}^2&= 0.96\mbar_0^2\TeV^2,
\ & \sumthr b \cr
m_{\etil_L}^2 + m_{\etil_R}^2+m_{\util_L}^2+m_{\util_R}^2 
- 1.70m^2_{\tilde g}&= 
-3.56\mbar_0^2\TeV^2, & \sumthr c \cr
m_{\util_R}^2+m_{\dtil_R}^2+m_{\util_L}^2+m_{\dtil_L}^2 
- 3.51m_{\tilde g}^2 &= 0.90\mbar_0^2\TeV^2, & \sumthr d \cr
m_A^2 - 2\sec 2\beta\left (m_{\tautil_1}^2 + m_{\tautil_2}^2
-2m_{\tau}^2\right) - 0.49m_{\tilde g}^2 &= 
1.05\mbar_0^2\TeV^2. \ & \sumthr e \cr}$$
Eqs.~\sumthr{c,d}\ above involve only the  first (or second) generations,
and  so the numerical results here are also independent of $\tan\beta$.
Thus these two sum rules hold for every column in Table~2, as is
easily verified. 

It is interesting to compare these sum rules with the corresponding 
ones in the Fayet-Iliopoulos  scenario described in our previous
paper\jjnew; essentially the distinction lies in the non-zero 
RHS in Eqs.~\sumthr{a-e}.

In conclusion, we have shown that within the MSSM  it is possible to
construct a  solution to the running equations for $m^2$, $M$ and $h$
that is  completely RG invariant, and leads to a phenomenologically
acceptable theory, resulting  in a distinctive spectrum with  sum rules
for the sparticle masses. Two sources of supersymmetry-breaking are
required, one corresponding to the gravitino mass (at around $m_0 =
40\TeV$) and another, related  to a $\Rcal$-symmetry, at around $|\mbar_0|
= 300-500\GeV$. The magnitude of the latter  suggests the idea of a
common origin for it, the $\mu_s$ term and the associated  $H_1 H_2$
soft term. A convincing demonstration of this  would considerably
enhance the attractiveness of this model.  It would also  be interesting
to consider variations on the same theme; forbidding the  Higgs $\mu_s$
term, or incorporating massive neutrinos, for example. 

\bigskip\centerline{{\bf Acknowledgements}}\nobreak

This work was supported in part by a Research Fellowship from the 
Leverhulme Trust. We thank the referee for helping us to clarify some issues.
\listrefs
\end